\documentclass[12pt,preprint]{aastex}
\usepackage{graphicx}
\usepackage[section]{placeins}

\newcommand\ha{\hbox{H$\alpha$}}
\newcommand\ic{$I_\mathrm{C}$}
\newcommand\rc{$R_\mathrm{C}$}
\newcommand\ks{$K_\mathrm{s}$}
\newcommand\mum{\hbox{$\mu$m}}
\newcommand\tm{\textit{2MASS\/}}
\newcommand\wise{\textit{AllWISE\/}}
\newcommand\spitzer{\textit{Spitzer\/}}

\newcommand\sdss{\textit{SDSS\/}}

\begin{document}

\title{New Candidate Eruptive Young Stars in Lynds~1340}
 \author{M. Kun\altaffilmark{1}, D. Apai\altaffilmark{2}, 
 J. O'Linger-Luscusk\altaffilmark{3}, A. Mo\'or\altaffilmark{1}, 
 B. Stecklum\altaffilmark{4}, E. Szegedi-Elek\altaffilmark{1}, G. Wolf-Chase\altaffilmark{5,6}}
 \email{kun@konkoly.hu}
 \altaffiltext{1}{Konkoly Observatory, H-1121 Budapest, Konkoly Thege \'ut  15--17, Hungary}
 \altaffiltext{2}{Department of Astronomy and Department of Planetary Sciences, The University of Arizona, 933 North Cherry Avenue, Tucson, AZ 85721, USA}
 \altaffiltext{3}{On leave from California Institute of Technology, 1200 E
California Ave, Pasadena, CA 91125}
 \altaffiltext{4}{Th\"uringer Landessternwarte Tautenburg, Sternwarte 5, 07778, Tautenburg, Germany}
 \altaffiltext{5}{Astronomy Department, Adler Planetarium, 1300 S. Lake Shore Drive, Chicago, IL 60605,USA}
 \altaffiltext{6}{Dept. of Astronomy \& Astrophysics, University of Chicago, 5640 S. Ellis Ave., Chicago, IL 60637, USA}

\date{Received 16 August 2014 / Accepted 3 October 2014}

\begin{abstract}
We report on the discovery of three candidate eruptive young stars, found during our comprehensive multi-wavelength study of the young stellar population of the dark cloud  L1340. These stars are as follows. (1) IRAS~02224+7227 (2MASS~02270555+7241167, HH\,487\,S) exhibited FUor-like spectrum in our low-resolution optical spectra. The available photometric data restrict its luminosity to 23\,$L_{\sun} < L_\mathrm{bol} < 59$\,$L_{\sun}$. (2) 2MASS~02263797+7304575, identified  as a classical T~Tauri star during our \ha\ survey, exhibited an EXor type brightening in 2005 November, at the time of the \sdss\ observations of the region. (3) 2MASS~02325605+7246055, a low-mass embedded young star, associated with a fan-shaped infrared nebula, underwent an outburst between the DSS1 and DSS2 surveys, leading to the appearance of a faint optical nebula. Our $[$\ion{S}{2}$]$ and \ha\ images, as well as the \spitzer\ IRAC 4.5\,\mum\ images revealed Herbig--Haro objects associated with this star. Our results suggest that amplitudes and time scales of outbursts do not necessarily correlate with the evolutionary stage of the stars.
\end{abstract}

\maketitle
                                                                                
\section{Introduction}
\label{Sect_1}

Eruptive young stellar objects (YSOs) are characterized by dramatically increased accretion from the circumstellar disk onto the star \citep{HK96}. In addition to the classical FU Orionis \citep[FUor;][]{Herbig77,HK96,Reipurth10} and EX Lupi \citep[EXor;][]{Herbig07,Herbig08,Lorenzetti12} type stars, observations of the past decades revealed several embedded eruptive young stars whose classification is uncertain, and differ from both classical types in amplitude and time scale \citep{Reipurth10,Kun11a,Kospal13}. Although the outbursts in each class are powered by enhanced accretion from the disk onto the star, the spectroscopic properties of FUors and EXors are radically different. EXors keep the major spectroscopic signatures of magnetospheric accretion during the outburst, whereas FUor spectra indicate star--disk interaction zones substantially transformed by the outburst. The powerful winds, associated with the high accretion rate, produce Herbig--Haro (HH) objects, which may thus be important tracers of the accretion history \citep{Reipurth10}.

Since a sizeable part of the stellar mass may build up during repeated outbursts \citep{VB06}, and the high accretion luminosity and associated outflows affect the structure and evolution of the protoplanetary disks, outbursting stars are key objects for understanding the formation of Sun-like stars and their planetary systems. Their extreme rarity and diversity, however, hinder the understanding of the origin and nature of the outbursts. The recent review by \citet{Audard14} lists 26 FUors and FUor-like objects, and 18 EXors can be found in the list of \citet{Lorenzetti12}. New discoveries and their subsequent monitoring are thus relevant to a better understanding of the phenomenon of episodic accretion.

The new candidate eruptive stars presented in this Letter are found in Lynds~1340, an isolated dark cloud at ({\it l,b\/}) = (130.1\degr,11.5\degr). The first large-scale studies of the region \citep{KOS94,KAY03,MMN03} suggested that L1340, located at 600\,pc from the Sun, has formed a few mid-B, A and early F type, and some two dozens of low-mass stars. We conducted a comprehensive multi-wavelength study of L1340, involving {\it Spitzer\/}, {\it WISE\/}, {\it 2MASS\/}, and {\it SDSS\/} photometric data, a search for \ha\ emission stars via slitless grism spectroscopy, long-slit optical spectroscopy, narrow-band optical and high angular resolution near infrared imaging observations (M. Kun et al. 2014, in preparation). This work led to a revised distance of $730\pm30$\,pc for L1340, and resulted in the discovery of some 250 candidate YSOs. Among them we found three new candidate eruptive young stars, which deserve special attention. One of them is IRAS~02224+7227, identified by \citet{KAY03} as the probable exciting source of HH~487, situated at 6.2\arcmin\ southwest of the star. The two others are 2MASS~02263797+7304575 and 2MASS~02325605+7246055, neither of them was mentioned before as a young star. We present the observational results which prove the eruptive nature of these objects. We adopt the revised distance of 730\,pc. 

\section{Data}
\label{Sect_data}

Lynds~1340 was observed by the \textit{Spitzer Space Telescope} using the Infrared Array Camera  \citep[IRAC;][]{Fazio2004} on 2009 March 16 and the Multiband Imaging Photometer for Spitzer  \citep[MIPS;][]{Rieke2004} on 2008 November 26 (Prog. ID: 50691, P.I. G. Fazio). The observations covered $\sim 1$~deg$^2$ in each band. The centers of the 3.6 and 5.8\,\mum\ images are slightly displaced from those of the 4.5 and 8\,\mum\ images, therefore the south-eastern part of the molecular cloud was not covered by the 4.5 and 8\,\mum\ maps. Moreover, the south-western part of L1340 is outside the field of view of the 24 and 70\,\mum\ images. We performed aperture photometry on individual saturation corrected basic calibrated data (CBCD) IRAC images, produced by the S18.18.0 pipeline at the Spitzer Science Center (SSC), using a 2-pixel aperture radius, and a sky annulus between 2 and 6 pixels. An additional array-dependent photometric correction and a pixel-phase correction \citep[see][]{Hora2008}, as well as an aperture correction were applied. The final photometry and its uncertainty were estimated as the average and rms of the individual flux densities measured in different CBCD frames.
For MIPS data we used MOPEX \citep{Makovoz2006}  to create mosaic maps from the 24\,\mum\ enhanced BCD and 70\,\mum\ BCD products (version S18.13.0) of SSC. Following \citet{Gordon2007} at 70\,\mum\  before mosaicing we made a column mean subtraction and a time filtering on the BCD frames. PSF photometry was used to determine the targets' fluxes in the mosaic maps. The resulted fluxes of the three targets are shown in Table\,\ref{Tab1}. The quoted uncertainties were computed as a  quadratic sum of the measurement errors and absolute calibration errors of 2\% for IRAC  \citep{Hora2008} and 4 and 7\%  for 24 and 70\,\mum\ MIPS  observations \citep{Engelbracht2007,Gordon2007}, respectively.

We obtained  low resolution optical spectra of the star coinciding with IRAS~02224+7227 on 2003 February 5 using CAFOS\footnote{\url{http://w3.caha.es/CAHA/Instruments/CAFOS/}} with the G--100 grism on the 2.2-m telescope of the Calar Alto Observatory, and on 2004 December 11 using FAST on the 1.5-m FLWO telescope \citep{Fabricant}. We reduced and analysed the spectra in IRAF.

High angular resolution {\it JHK\/} images, centered on the same star, were obtained on 2002 October 24 using the near-infrared camera Omega-Cass\footnote{\url{http://www.caha.es/CAHA/Instruments/OCASS/ocass.html}}, mounted on the 3.5-m telescope of the Calar Alto Observatory. 
The plate scale was 0.1\arcsec/pixel. The target was observed at four dithering positions, and the observations consisted of two dither cycles. The total integration time  was 480\,s in  each filter. 
We reduced and analysed the data in IRAF. Following the flat-field correction and bad pixel removal the sky frame for each cycle was obtained by taking the minimum of the images at different dithering positions. This sky frame was subtracted from each individual image of a given cycle, and the frames from a single cycle were combined into a mosaic image. Aperture photometry was performed on the reduced images. The instrumental magnitudes were transformed into the {\it JHK}$_\mathrm{s}$ system by using the \tm\ magnitudes of field stars within the field of view. The resulting magnitudes are listed in Table~\ref{Tab1}.

We performed a new search for \ha\ emission stars in L1340  using the Wide Field Grism Spectrograph~2 ({\it WFGS2}) installed on the University of Hawaii 2.2-meter telescope.
The instrument setup and data reduction procedure were same as described in detail in \citet{Szeg13}. We observed 2MASS~02263797+7304575 on 2011 October 16, and detected a \ha\ emission with $EW(\ha)=-80$~\AA\ in its spectrum.

The \ks\ magnitude of 2MASS~02325605+7246055 was measured on the images obtained on 2010 October 18, during the monitoring program of V1180~Cas \citep{Kun11a}, using the MAGIC camera on the 2.2-m telescope of the Calar Alto Observatory.

Narrow band images through $[$\ion{S}{2}$]$ and \ha\ filters, as well as broad \textit{R}-band images containing the environment of 2MASS~02325605+7246055, were obtained with the Schmidt telescope of the Th\"uringer Landessternwarte (TLS), Tautenburg in May, June and September 2011. The frames obtained through the same filter were coadded, leading to integration times of 0.2, 3.0, and 4.0 hours in the $R$, \ha, and $[$\ion{S}{2}$]$, respectively. Spectra of the nebula and the two brightest HH knots were obtained using the TLS medium-resolution Nasmyth spectrograph ($R\sim700$) in 2011 November. The total integration time was 2 hours per object.

{\it BVR$_\mathrm{C}$I$_\mathrm{C}$\/} photometric observations of IRAS~02224+7227 were performed with the 1-m Ritchey--Chretien--Coude (RCC) telescope of the Konkoly Observatory at three epochs between 2001 and 2011. We measured the {\it R}$_\mathrm{C}$ and {\it I}$_\mathrm{C}$ magnitudes of IRAS~02224+7227 and 2MASS~02263797+7304575 at several epochs between 2011 January and 2014 June on the images collected with the wide-field camera on the Schmidt telescope of the Konkoly Observatory to monitor the light variations of V1180~Cas \citep{Kun11a}. The results of the photometry are listed in the machine-readable version of Table~\ref{Tab1}. 

L1340 is situated within Stripe~1260 of the {\it SEGUE\/} survey \citep{Yanny2009}, thus its whole area was observed in the {\it ugriz\/} bands in 2005 November--December. Each target star has high-quality 3.4, 4.6, 12, and 22 micron fluxes in the \wise\ data base. These data offer useful pieces of information on their spectral energy distribution (SED) and long-term photometric behavior. 

\section{Results and Discussion}

\subsection{IRAS~02224+7227: a FUor-like Star}

\begin{figure*}
\centerline{\includegraphics[width=12cm]{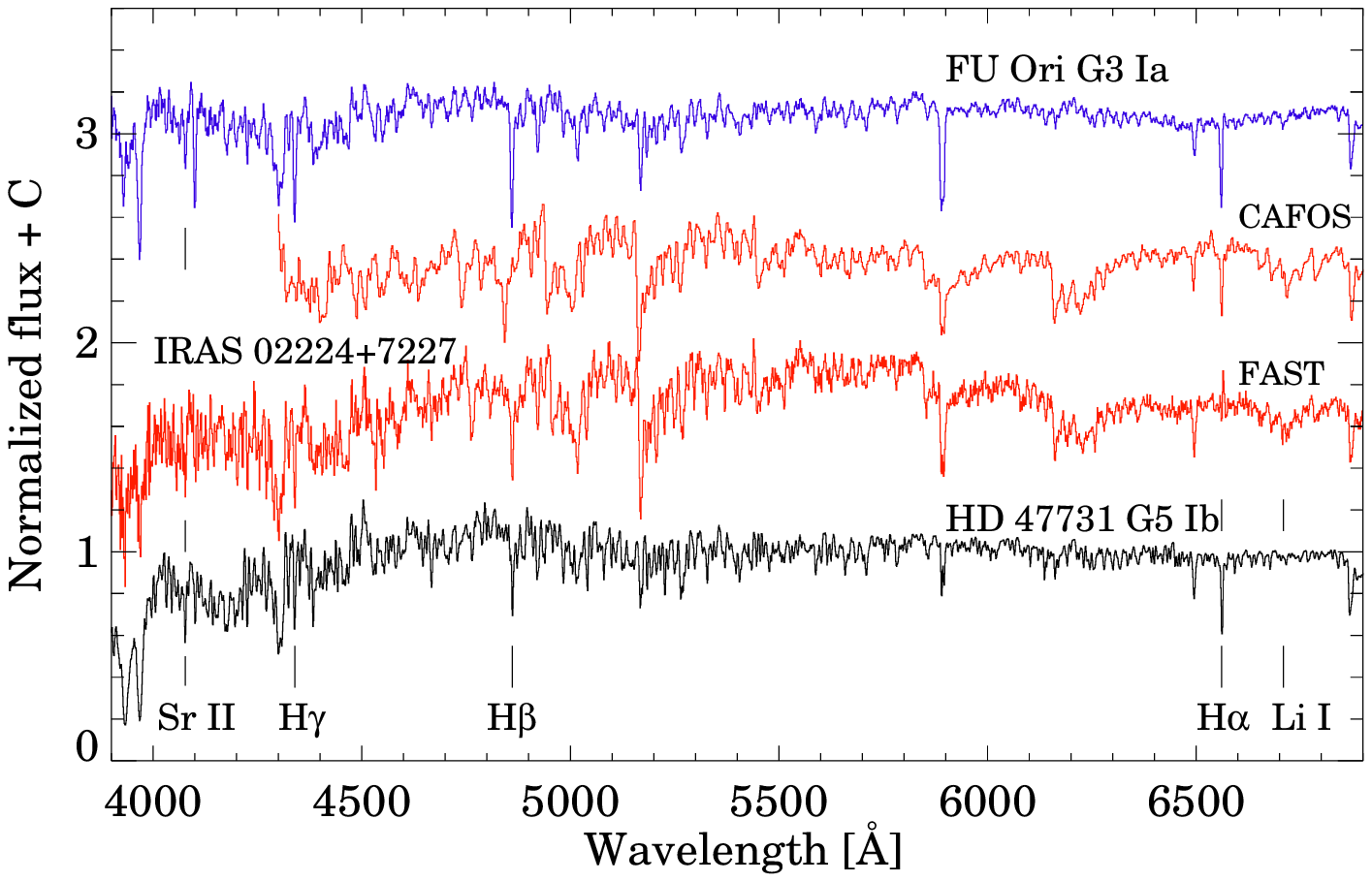}}
\centerline{\includegraphics[width=8cm]{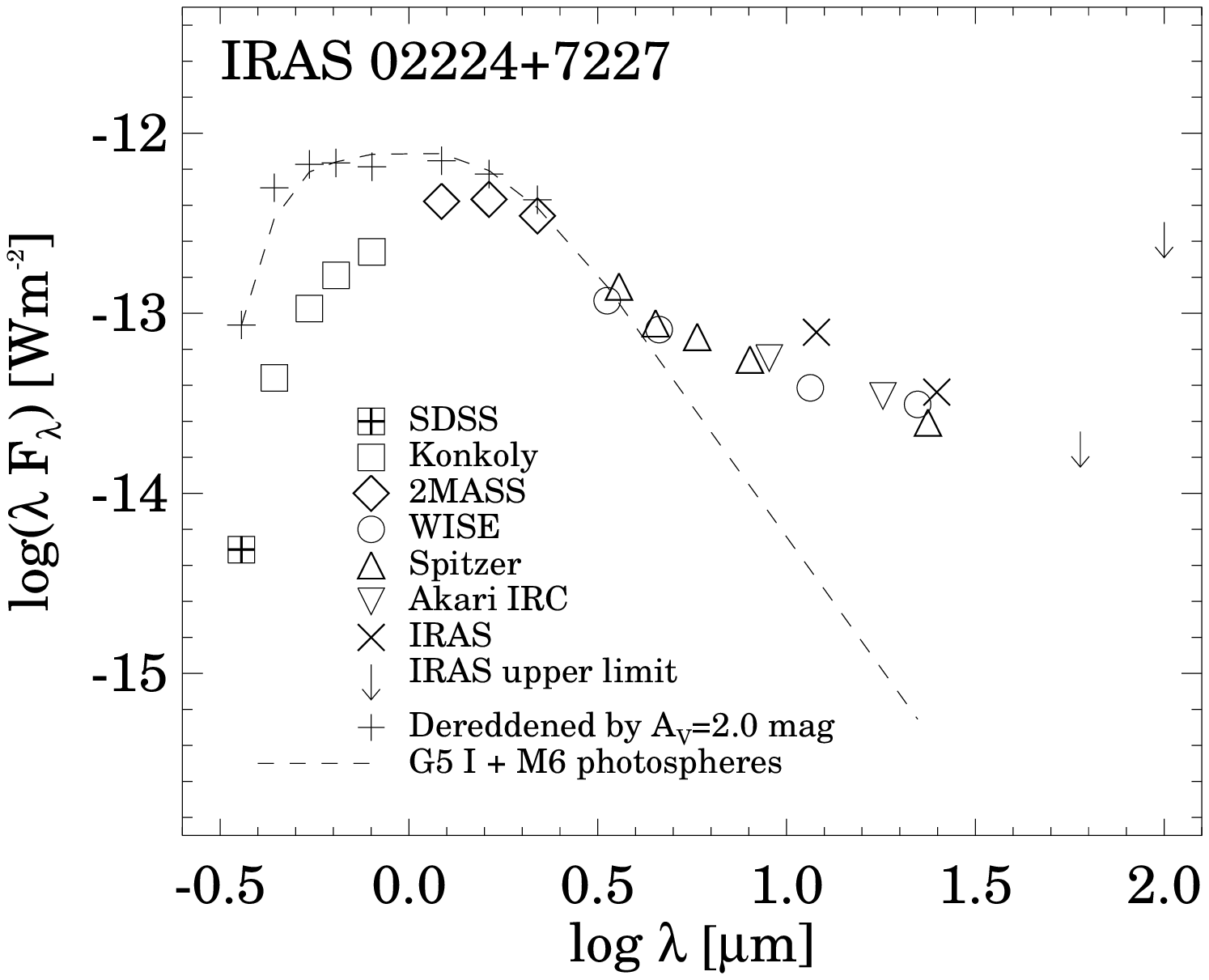}
\includegraphics[width=5.5cm]{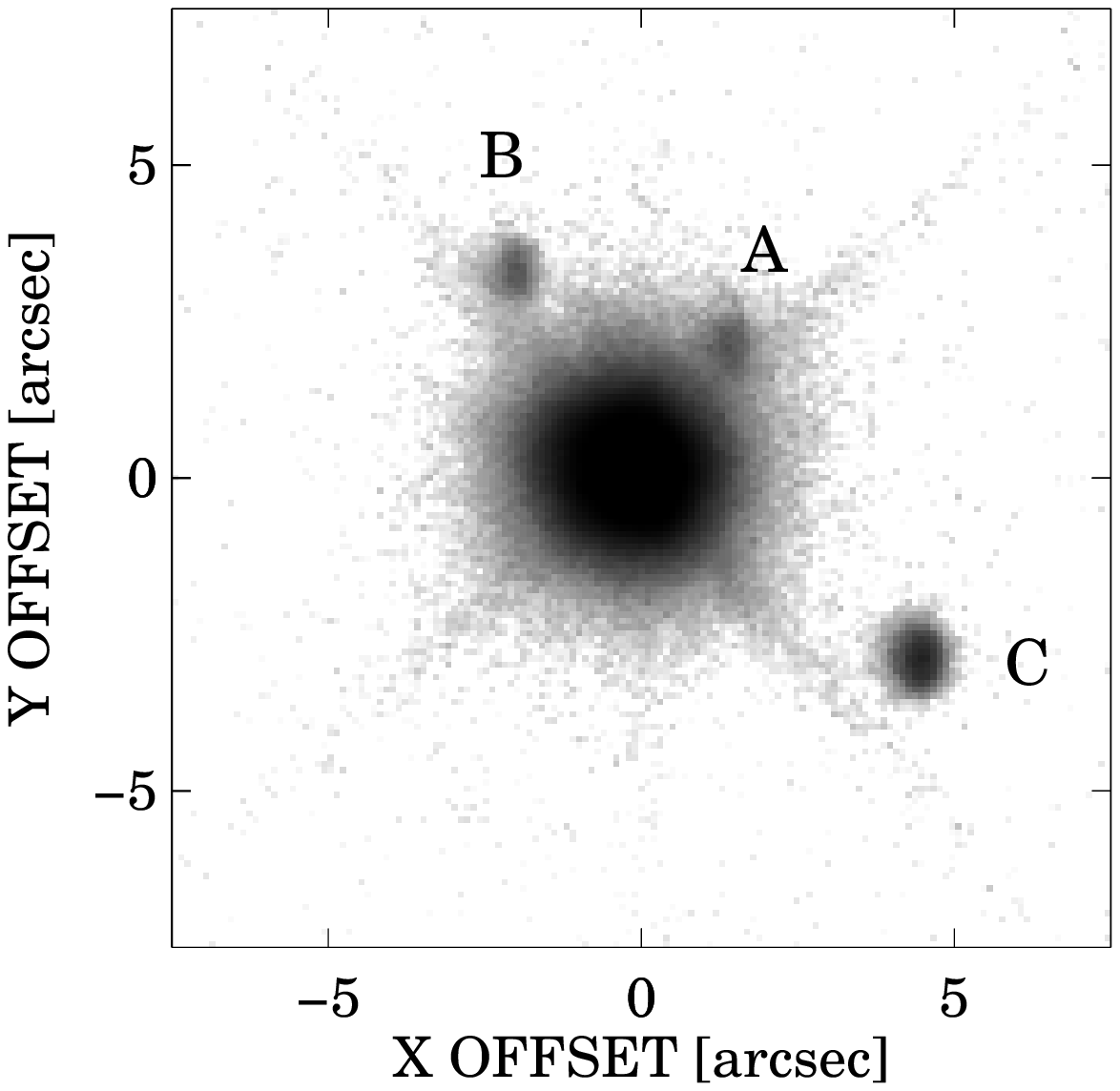}} \vskip 6mm
\caption{(1) Spectra of IRAS 02224+7227 (red), compared with that of FU Orionis (blue) and the spectrum of the G5~Ib type supergiant HD~47731 (black) \citep{LeBorgne}. (2) SED of IRAS 02224+7227. (3) Visual companions of the same star in the Omega-Cass \textit{K\/}-band image.}
\label{Fig1}
\end{figure*}

IRAS~02224+7227 (2MASS~02270555+7241167, HH\,487\,S) is situated near the south-western edge of L1340. The upper panel of Fig.~\ref{Fig1} shows its FAST and CAFOS spectra, together with that of FU~Ori, found in the FAST Public Archive\footnote{\url{http://tdc-www.harvard.edu/cgi-bin/arc/fsearch}}, and with the spectrum of the G5-type supergiant star HD~47731, found in \citeauthor{LeBorgne}'s \citeyearpar{LeBorgne} spectrum library. The hydrogen and metallic absorption spectra, characteristic of G-type stars, the \ion{Sr}{2} line at 4077\,\AA, indicative of high luminosity class, and the youth-indicator \ion{Li}{1} line at 6707\,\AA\ can clearly be identified, suggesting the FU~Ori nature of this star. Comparison of the strength of the H$\gamma$, H$\beta$  lines and the G-band with those in spectroscopic standards \citep{JHC84,LeBorgne} suggests G4--G5 spectral type. Weak \ha\ emission can be seen in the FAST spectrum, while the \ha\ line is in absorption in the CAFOS spectrum. We estimated a foreground extinction of $A_\mathrm{V} \approx 2.0$~mag from the G5 spectral type, and the ({\it R\/}$_\mathrm{C}-${\it I\/}$_\mathrm{C}$ ) color index, adopting an intrinsic color index of ({\it R\/}$_\mathrm{C}-${\it I\/}$_\mathrm{C})_0= 0.37$ \citep{HK96}.

The second panel of Fig.~\ref{Fig1} shows the SED of the star, constructed from \sdss\ {\it u\/}, \citep[transformed into Johnson {\it U\/} following][]{Jordi06}, \spitzer, \tm, \wise, and our own {\it BVR$_\mathrm{C}$I$_\mathrm{C}$\/} data. IRAS~02224+7227 lies outside the field of view of the 70\,\mum\ MIPS image, and only flux upper limits are available for the IRAS 60 and 100\,\mum\ bands. The shape of the dereddened SED in the 0.36--3.6\,\mum\ region can be satisfactorily approximated with the sum of a G5I and an M6 type photosphere, contributing nearly equally to the total flux at 2\,\mum, and supporting that FUors exhibit wavelength-dependent spectral types \citep{HK96}. Integrating the dereddened SED over the 0.36--24\,\mum\ wavelength interval, and assuming a distance of 730~pc we obtain a luminosity of $L(0.36-24) \approx 23$\,$L_{\sun}$. 
Including the {\it IRAS\/} 60\,\mum\ and 100\,\mum\ flux upper limits we obtain $L(0.36-100) < 59$\,$L_{\sun}$. Although most of the known FUors have $L_\mathrm{bol} > 100 L_{\sun}$, both values are within the range of FUor luminosities ($7 \la L_\mathrm{bol}/L_{\sun} \la 800$) compiled by \citet{Audard14}.

The third panel of Fig.~\ref{Fig1} shows the central part of the Omega-Cass {\it K\/}-band image. IRAS~02224+7227 has three wide companions. Table~\ref{Tab2} lists the separations from the central star, position angles (from the North towards the East), and {\it JHK$_\mathrm{s}$} magnitudes of the three stars. Each companion has normal stellar colors in the near-infrared bands. The angular separations correspond to 1750--4000~AU at 730\,pc, comparable with the typical size of protostellar envelopes. Spectroscopy and/or L/M-band photometry are required to decide whether these stars are physically related to each other or not.

Since the brightness of the star in the POSS\,1 image, recorded on 1954 September 27, is similar to the more recent ones shown in Table~\ref{Tab1}, IRAS~02224+7227 has been in outburst for at least sixty years. Similarly to other FUor-like stars whose outburst dates are unknown, we observe in this star an evolved phase of outburst. If HH~487 was created by a previous outburst of the star and we assume a typical HH-object space velocity 200~km\,s$^{-1}$, then its angular distance of 6.2\arcmin, corresponding to 1.3\,pc without accounting for the unknown inclination, suggests that a previous outburst might have happened some 6500 years ago. This interval corresponds to statistical estimates on the average time span between outbursts \citep{Scholz13}. The present outburst thus might have occurred in a disk shaped by previous outburst(s). The Class~II SED, absence of bright reflection nebulosity, and the relatively low luminosity may also suggest an `old' FUor. Our photometric measurements indicate no trend in the optical magnitudes between 2001 and 2014 (Table~\ref{Tab1}).

\subsection{2MASS 02263797+7304575: a possible EXor}

\begin{figure*}
\centerline{\includegraphics[width=8cm]{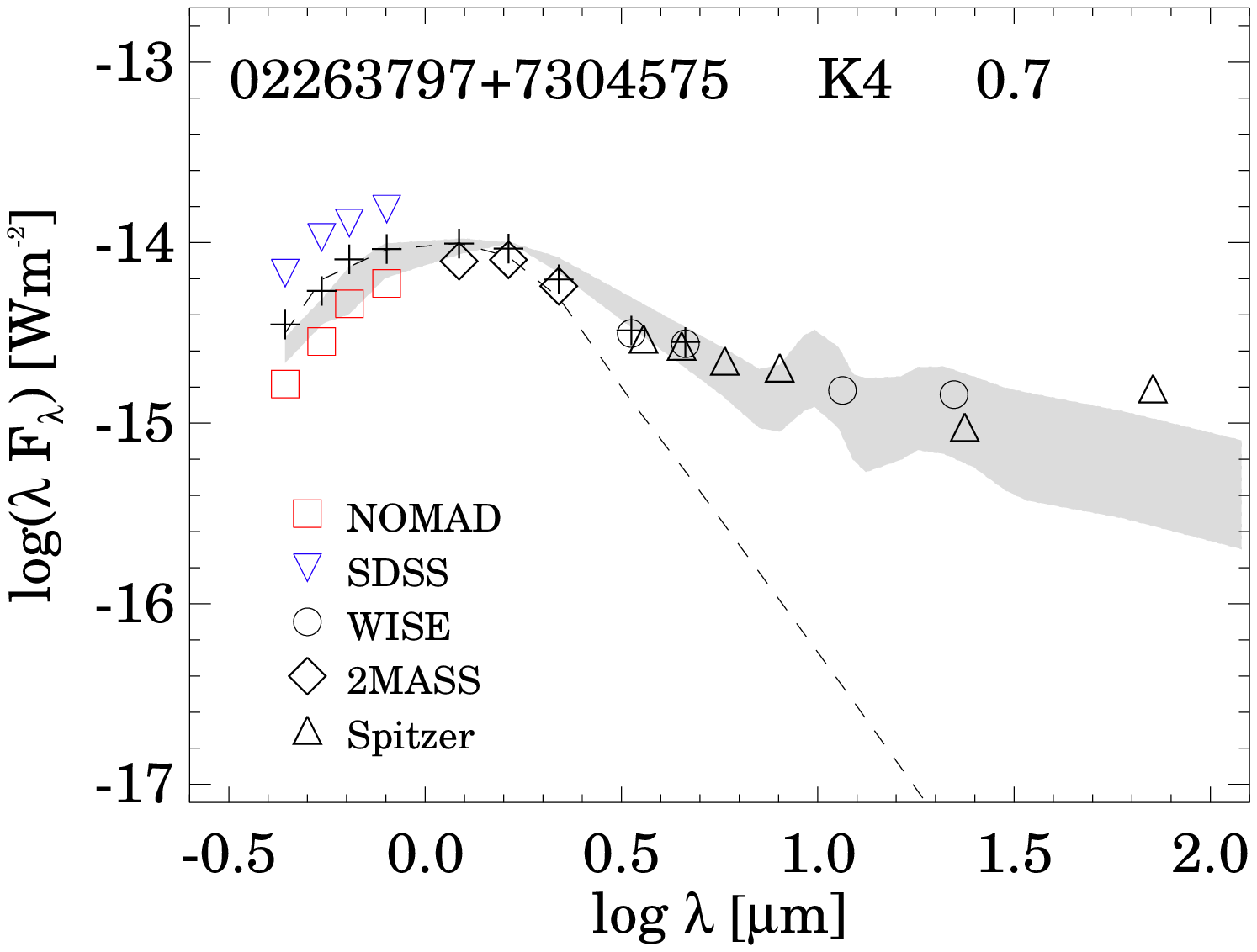}
\includegraphics[width=6cm]{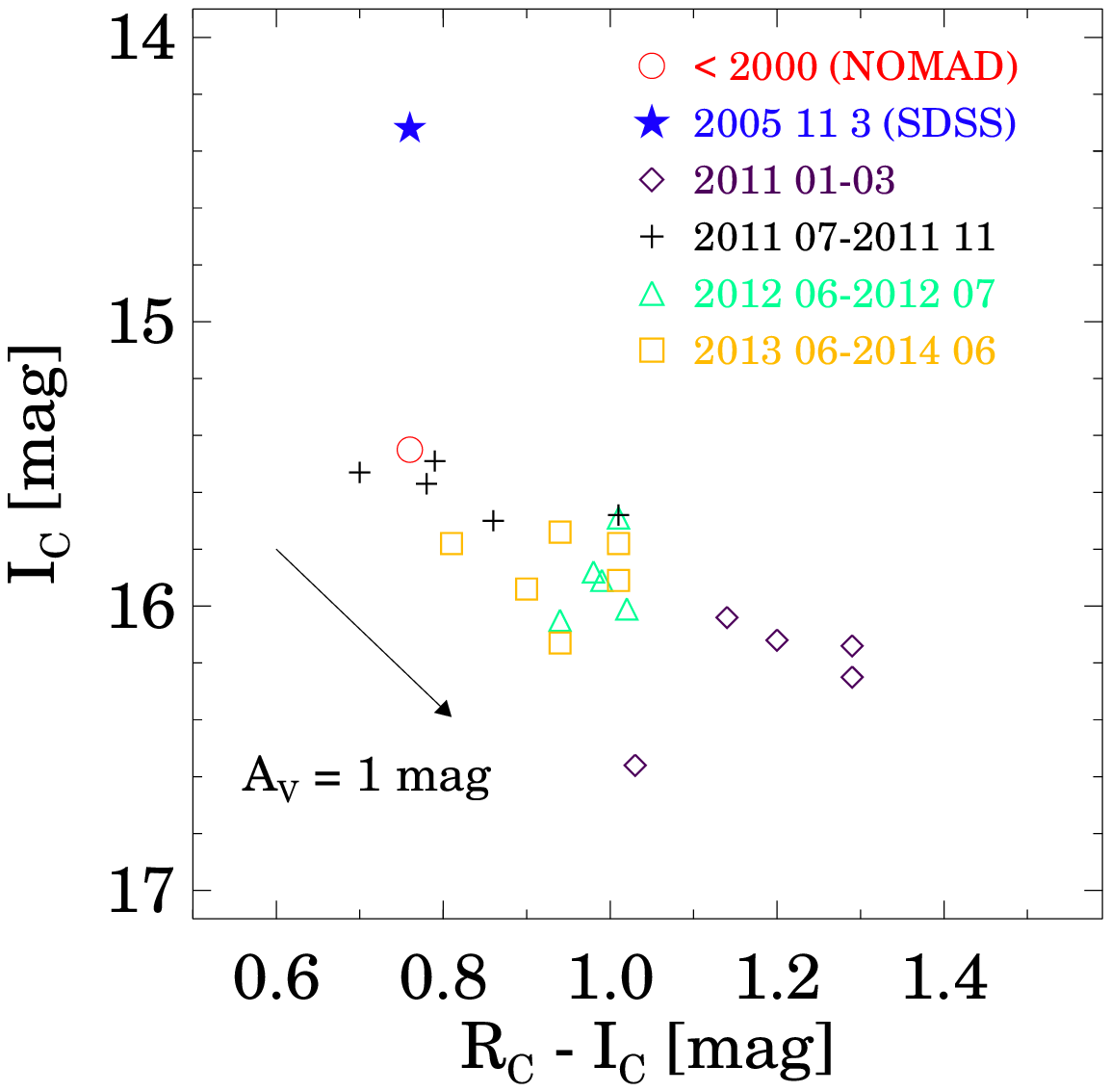}}
\caption{{\it Left\/}: Spectral energy distribution of 2MASS 02263797+7304575. Crosses indicate the low-state fluxes, corrected for an extinction of $A_\mathrm{V} = 0.7$\,mag, and the dashed line is a K4 type photosphere fitted to the dereddened fluxes. The gray band indicates the median SED of the Taurus pre-main sequence stars, fitted to the dereddened SED at 1.25\,\mum. {\it Right\/}: {\it R}$_\mathrm{C}$ vs. {\it R}$_\mathrm{C}-${\it I}$_\mathrm{C}$ color--magnitude diagram of the same star, containing all available data.}
\label{Fig2}
\end{figure*}

We identified this star as a candidate classical T~Tauri star (CTTS) based on the strong \ha\ emission ($EW(\ha)=-80$~\AA) in its spectrum. The shape of its SED, constructed from the {\it NOMAD\/} {\it BVRI\/}, \sdss, \tm, \wise, and \spitzer\ data, and plotted in Fig.~\ref{Fig2} (left), together with the band of typical SEDs of the Taurus pre-main sequence stars \citep{DAlessio}, confirms the CTTS nature. The \sdss\ {\it griz\/} magnitudes, measured on 2005 November 3, were transformed into the {\it BVR$_\mathrm{C}$I$_\mathrm{C}$\/} system \citep{Ivezic07} to compare them with other available photometric data. It is apparent that while the optical magnitudes of the {\it NOMAD\/} catalogue, the \tm, \spitzer, and \wise\ data, measured at various epochs, smoothly delineate a Class~II SED, the \sdss\ magnitudes stand apart, suggesting that the star was unusually bright over the optical spectrum on 2005 November 3. We estimated the spectral type, extinction, and luminosity of the star by comparing the short-wavelength (\textit{BVRIJ}) side of the low-state SED with a grid of reddened photospheres, using the pre-main sequence colors and bolometric corrections tabulated by \citet{Pecaut2013} and the extinction law of \citet{CCM89}, as well as the $A_\mathrm{V} \ge 0.5$\,mag  \citep[foreground extinction toward L1340,][]{KWT03} restriction. The dashed line shows the best estimate, the SED of a K4-type photosphere, fitted to the {\it NOMAD\/} and \tm~{\it J\/} magnitudes, dereddened by $A_\mathrm{V}=0.7$\,mag. Its photospheric luminosity is 0.28\,$L_{\sun}$ at a distance of 730\,pc. The total luminosity of the system, derived by integrating the dereddened SED and extrapolating the contribution of the spectral regions beyond 70\,\mum\ using the method by \citet{Chavarria}, is $L_\mathrm{bol} \approx 0.39$\,$L_{\sun}$, suggesting $L_\mathrm{disk} / L_\mathrm{star} \approx 0.39$,
higher than the upper limit ($\sim0.2$) for passive irradiated disks \citep{KH87}.

The {\it R}$_\mathrm{C}$ vs. {\it R}$_\mathrm{C}-${\it I}$_\mathrm{C}$ color--magnitude diagram is plotted in Fig.~\ref{Fig2} (right). It can be seen that, with the exception of the \sdss\ point, the star's  {\it R}$_\mathrm{C}$ magnitudes stay in the $15.4 <${\it R}$_\mathrm{C} < 16.3$~mag interval. The amplitude of the light variations within an observing season is about 0.5\,mag, and the average brightness changes from season to season. The distribution of the points suggests variable circumstellar extinction. The single bright point indicates a burst-like event with an amplitude of 1.5--2.0~mag. Similar outbursts are supposed to frequently occur in CTTSs, but due to their short duration are hard to catch. Detection of such events may be helpful in looking for the specialities of EXor disks, distinguishing them from those of less violent pre-main sequence stars. Such specialities may be
the high  $L_\mathrm{disk} / L_\mathrm{star}$, and the flattening of the SED  between 24 and 70\,\mum, unlike typical T~Tauri SEDs, and indicative of a remnant infalling envelope \citep{Calvet94}. A similar flattening can be seen in the SED of prototype EXor, EX Lupi \citep{Sipos}, and 
the candidate EXors identified by \citet{Giannini} also exhibit flat SEDs near the ClassI/ClassII boundary, indicating that eruptive events occur at the earliest evolutionary phases of the protoplanetary disks.

\subsection{2MASS~02325605+7246055: an Outbursting Protostellar Object?}

\begin{figure*}
\centerline{\includegraphics[width=6cm]{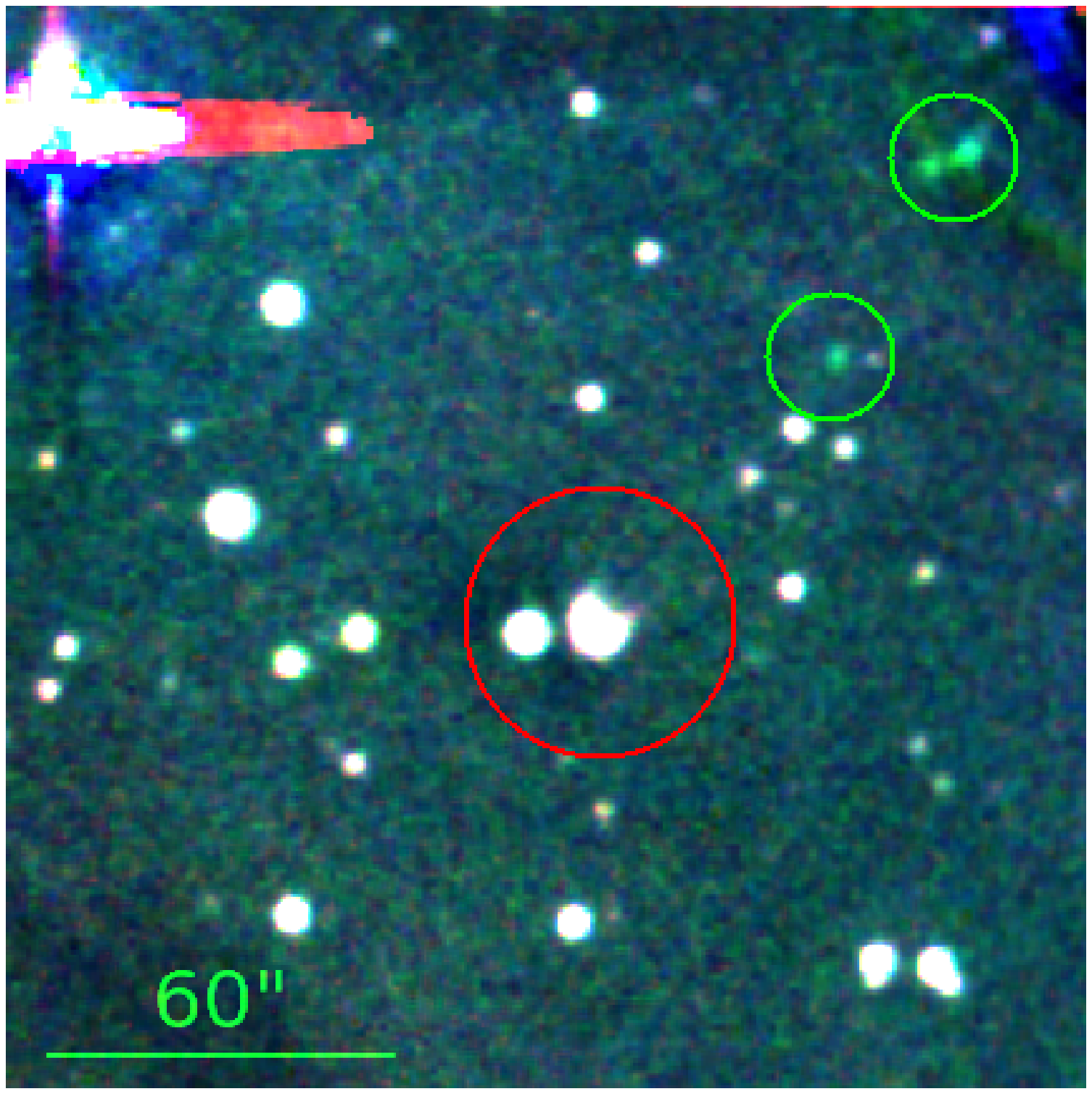} \hskip 15mm
\includegraphics[width=6cm]{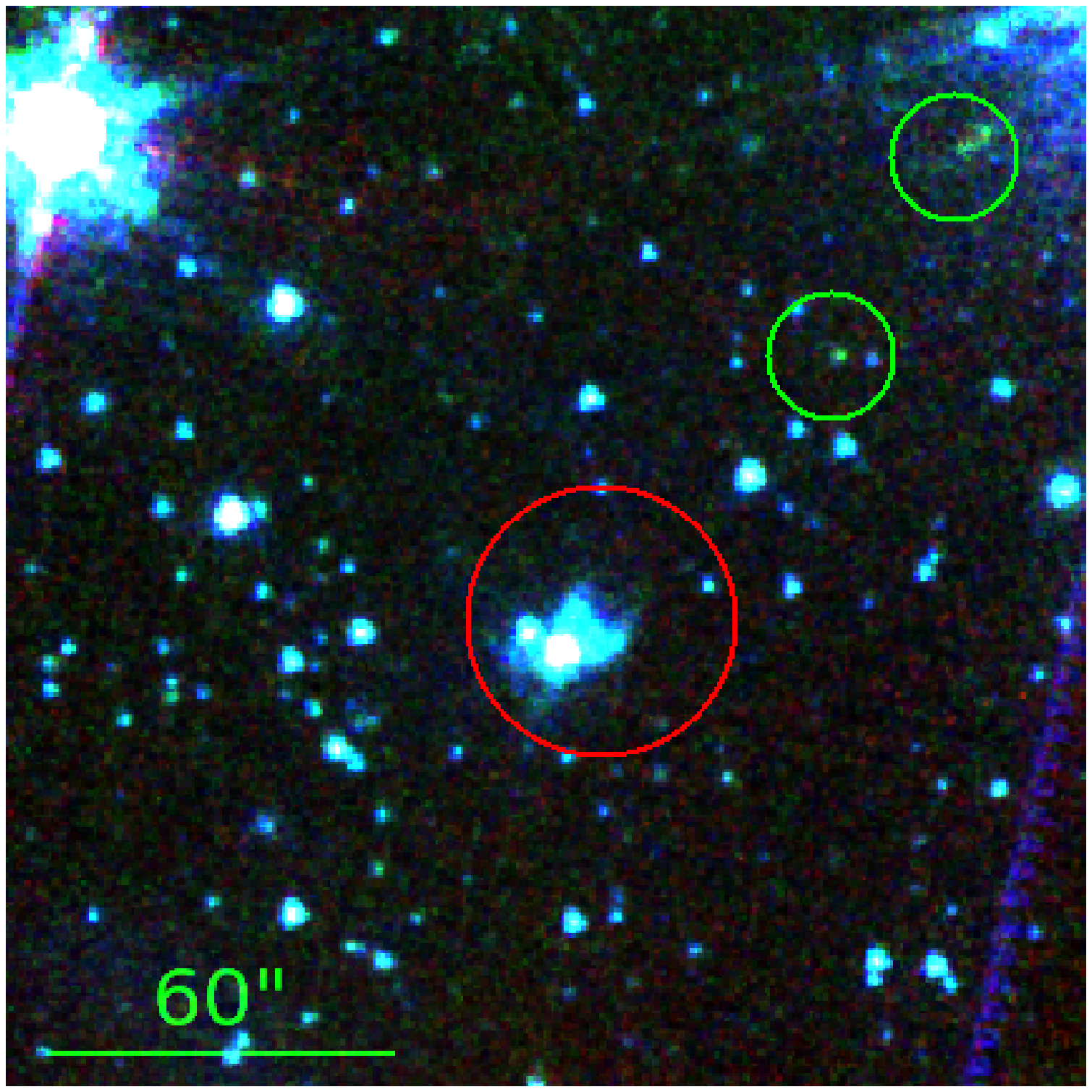}}
\vskip 5mm
\centerline{\includegraphics[width=8cm]{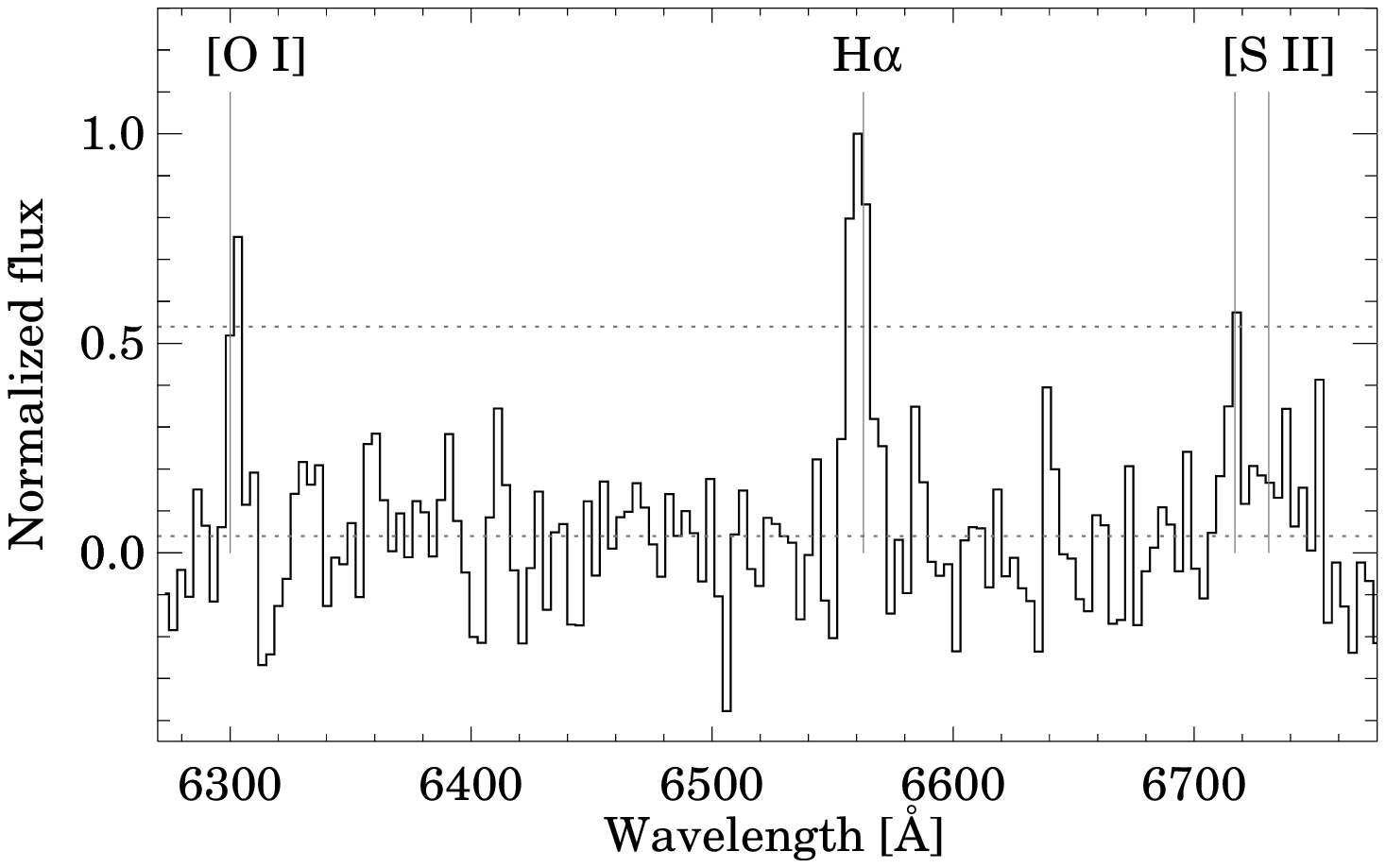} \hskip 5mm
\includegraphics[width=8cm]{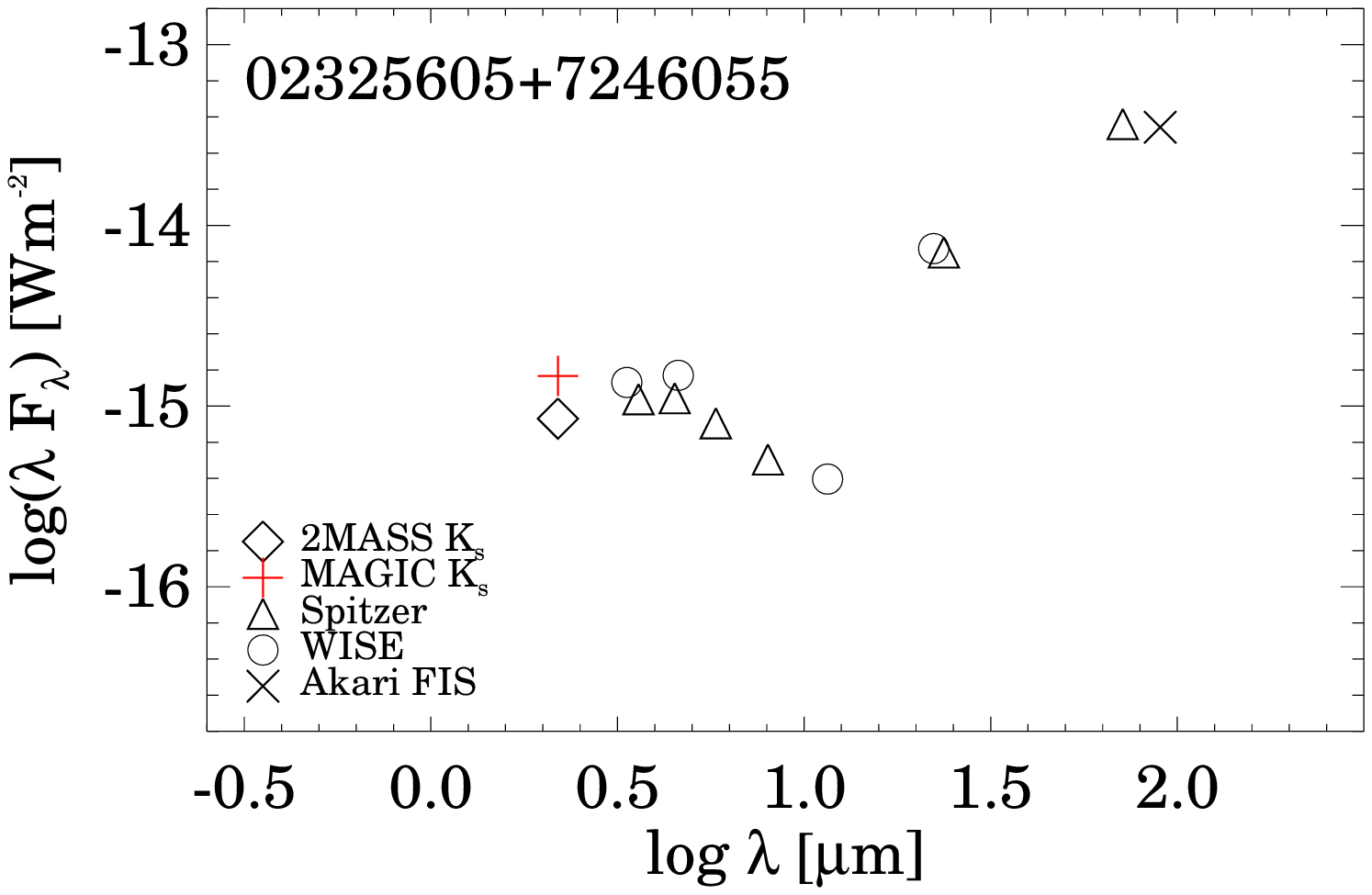}}
\caption{{\it Top left\/}: Three-color image, composed of \rc\ (red), \ha\ (green), and [\ion{S}{2}] (blue) images of  the environment of 2MASS~02325605+7246055 (located within the red circle), obtained with the Tautenburg Schmidt telescope. Green circles encompass the HH objects. {\it Top right\/}: IRAC 3.6 (blue), 4.5 (green), and 5.8 (red) composite image of the same area.
 {\it Bottom left\/}: Part of the spectrum of the \textit{NW} HH knot. The dotted horizontal lines show the mean and 3-$\sigma$ levels. The vertical lines indicate rest wavelengths.  
{\it Right\/}: SED of 2MASS~02325605+7246055.}
\label{Fig3}
\end{figure*}

This star is invisible in the DSS\,1 red image. A faint bow-shaped nebula appears near its position in the DSS\,2 red image, indicating that the outburst of an embedded star opened a cavity in its circumstellar envelope between 1954 September 27 and 1994 January 3. The nebula is also visible in our {\it R}$_\mathrm{C}$ and {\it I}$_\mathrm{C}$ images, obtained since 2001 with various instruments of the Tautenburg and Konkoly Observatories. The shortest wavelength where the star appears is the K-band. Our \ks\ image obtained in 2010, as well as the \spitzer\ 3.6 and 4.5\,\mum\ images (Fig.~\ref{Fig3}, upper right panel) show a small fan-shaped nebula next to the star at the northwestern side, similar to those found near several embedded eruptive stars (e.g. RNO~125 associated with PV~Cep \citep{Kun11b}, and McNeil's Nebula with V1647~Ori \citep{Briceno04}). The \ks\ magnitude of the star, measured in 2010, was some 0.5 mag brighter than the \tm\ \ks\ (Table~\ref{Tab1}), suggesting a years time-scale brightening of the star. Our \ha\ and $[$\ion{S}{2}$]$ images (Fig.~\ref{Fig3}, upper left) show two faint HH knots, located at 71\arcsec\ (SE) and 112\arcsec\ (NW) to the northwest of the star, at (RA, Dec)(SE)=($2^\mathrm{h}32^\mathrm{m}45.3^\mathrm{s}; +72\degr46\arcmin56\arcsec$), and (RA, Dec)(NW)=($2^\mathrm{h}32^\mathrm{m}40.2^\mathrm{s}; +72\degr47\arcmin32\arcsec$), respectively. The HH knots are associated with shocked H$_2$ emission seen in \textit{IRAC\/} 4.5-micron image. While the spectrum of the \textit{SE} HH knot has low signal-to-noise ratio, the \textit{NW} knot exhibits \ha, $[$\ion{O}{1}$]$, and $[$\ion{S}{2}$]$ emission lines (Fig.~\ref{Fig3}, lower left panel). The radial velocity of the \ha\ is $v_\mathrm{LSR} = -85\pm30$\,km\,s$^{-1}$, consistent with the cometary morphology of the object in the sense that we see the scattered light from the blueshifted outflow lobe. The monopolar morphology in the optical/IR implies a significant inclination. The spectrum of the object shows a faint continuum but lacks \ha\ emission. This indicates that at present accretion is weak or even absent which points to the lack/replenishment of the inner disk, most probably depleted during the outburst. The SED, plotted in the lower right panel of Fig.~\ref{Fig3}, suggests an embedded object with deep silicate absorption at 10\,\mum. Its bolometric luminosity, estimated from the 70\,\mum\ flux following the method of \citet{Dunham}, is $L_\mathrm{bol} \approx 1.2$\,$L_{\sun}$ at 730\,pc.  
The lack of \ha\ emission from the spectrum and the low bolometric luminosity suggest the present low-accretion state of 2MASS~02325605+7246055. The remnant of its recent outburst, the reflection nebula which appeared between 1954 and 1995, suggests an EXor-like event.

\section{Conclusions}

We identified three candidate eruptive young stars associated with the molecular cloud L1340, whose YSO population consists of some 250 members (Kun et al. 2014, in prep.). Together with the previously identified V1180~Cas \citep{Kun11a}, some 1.6\% of the total YSO poulation belong to eruptive classes. 
The observed properties of the FUor-like IRAS~02224+7227 suggest a late stage of its episodically accreting phase of evolution. The EXor candidate 2MASS~02263757+7304575 exhibits a Class~II SED, but its high 70\,\mum\ flux is indicative of a remnant infalling envelope. The nebulosity beside the Class~I protostar 2MASS~02325605+7246055 is a signature of a recent EXor-like outburst. These results indicate that amplitudes and time scales of YSO outbursts do not strictly correlate with the evolutionary stage of the star. The divergent observed properties of the eruptive young stars \citep[e.g.][]{Quanz07,Lorenzetti12} point to objects at various episodes of their consecutive outbursts.

\begin{acknowledgements}
We are grateful to G\'abor F\H{u}r\'esz for obtaining the FAST spectrum of IRAS~02224+7227.
 This work makes use of observations made with the \textit{Spitzer Space Telescope}, which is operated by the Jet Propulsion Laboratory, California Institute of Technology under a contract with NASA. Our results are partly based on observations collected at the Centro Astron\'omico Hispano Alem\'an (CAHA) at Calar Alto, operated jointly by the Max-Planck Institut f\"ur Astronomie and the Instituto de Astrof\'{\i}sica de Andaluc\'{\i}a (CSIC). This research utilized observations with the $2.2$-m telescope of the University of Hawaii and we thank Colin Aspin and Bo reipurth for their interest and support. Our research has benefited from the VizieR catalogue access tool, CDS, Strasbourg, France. Financial support from the Hungarian OTKA grant K81966 is acknowledged.
\end{acknowledgements}

\clearpage

\clearpage

\clearpage

\begin{deluxetable}{clrrrl} 
\tabletypesize{\scriptsize}
\tablecolumns{6} 
\tablewidth{0pc} 
\tablecaption{Photometric data of the target stars.\tablenotemark{*} \label{Tab1}}
\tablehead{ 
\colhead{Date} & \colhead{Band} & \colhead{2MASS\,02263797} & \colhead{IRAS\,02224+7227} & \colhead{2MASS\,02325605} & \colhead{Source\tablenotemark{a}} \\ 
\colhead{yyyymmdd}}
\startdata
19991028  & {\it J\/} (mag)      & 14.211\,(0.031)  &  9.901\,(0.023)  & \nodata & \tm\ \\
19991028  & {\it H\/} (mag)      & 13.403\,(0.037)  &  9.084\,(0.051)  & \nodata & \tm\  \\
19991028  & \ks (mag)            & 13.034\,(0.028)  &  8.579\,(0.020)  & 15.104\,(0.163) & \tm\ \\
20011014  & {\it B\/} (mag)      &  \nodata         & 14.656\,(0.060)  & \nodata &  Konkoly RCC \\ 
20011014  & {\it V\/} (mag)      &  \nodata         & 13.231\,(0.050)  & \nodata &  Konkoly RCC \\   
20011014  & \rc\ (mag)           &  \nodata         & 12.317\,(0.050)  & \nodata &  Konkoly RCC \\   
20011014  & \ic\ (mag)           &  \nodata         & 11.455\,(0.050)  & \nodata &  Konkoly RCC \\  
20021024  & {\it J\/} (mag)      &  \nodata         &  9.990\,(0.054)  & \nodata &  CA\,3.5-m/Omega-Cass \\
20021024  & {\it H\/} (mag)      &  \nodata         &  9.171\,(0.051)  & \nodata &  CA\,3.5-m/Omega-Cass \\
20021024  & \ks (mag)            &  \nodata         &  8.694\,(0.020)  &  \nodata & CA\,3.5-m/Omega-Cass \\
\enddata	
\tablenotetext{*}{The whole table is available only on-line as a machine-readable table.}
\tablenotetext{a}{Telescope/instrument or data base.}
\end{deluxetable}

\begin{deluxetable}{cccccr} 
\tabletypesize{\scriptsize}
\tablecolumns{6} 
\tablewidth{0pc} 
\tablecaption{Visual companions of IRAS 02224+7227 \label{Tab2}}
\tablehead{ 
\colhead{Star} & \colhead{{\it J}} & \colhead{{\it H}} & \colhead{\ks} & \colhead{Sep.} & \colhead{P.A.} \\
 & \colhead{(mag)} &  \colhead{(mag)} & \colhead{(mag)} & \colhead{(\arcsec)} & \colhead{(\degr)} } 
\startdata
 A & 15.82 & 15.30  & 14.96 & 2.4 & 324.0 \\
 B & 15.69 & 15.22  & 15.03 & 3.8 & 32.3 \\
 C & 14.98 & 14.32  & 14.01 & 5.4 & 236.4 \\    
\enddata  
\end{deluxetable}

\end{document}